# A multimodal approach for tracing lateralization along the olfactory pathway in the honeybee through electrophysiological recordings, morpho-functional imaging, and behavioural studies.

Albrecht Haase[1,*], Elisa Rigosi[2,3], Elisa Frasnelli[4], Federica Trona[3], Francesco Tessarolo[1], Claudio Vinegoni[5], Gianfranco Anfora[3], Giorgio Vallortigara[2], Renzo Antolini[1]

[1]Physics Department and BIOtech research centre, University of Trento, Via Sommarive 14, 38050 Povo (TN), Italy

[2]CIMeC, Centre for Mind/Brain Sciences, University of Trento, Corso Bettini 31, 38068 Rovereto, Italy

[3] Sustainable Agro-ecosystems and Bioresources Department, IASMA Research and Innovation Centre, Fondazione Edmund Mach
Via E. Mach 1, 38010 San Michele all'Adige, (TN), Italy

[4]Konrad Lorenz Institute for Evolution and Cognition Research, Adolf Lorenz Gasse 2, A-3422 Altenberg, Austria

[5]Center for Systems Biology and Center for Molecular Imaging Research, Massachusetts General Hospital, Harvard Medical School, 185 Cambridge Street, Boston, Massachusetts 02114, USA

* Tel.: +39 0461 28 3650

Fax: +39 0461 28 3659

E-mail: albrecht.haase@unitn.it

**Abstract**

Recent studies have revealed asymmetries between the left and right sides of the brain in invertebrate species. Here we present a review of a series of recent studies from our labs, aimed at tracing asymmetries at different stages along the honeybee's (*Apis mellifera*) olfactory pathway. These include estimates of the number of sensilla present on the two antennae, obtained by scanning electron microscopy, as well as electroantennography recordings of the left and right antennal responses to odorants. We describe investigative studies of the antennal lobes, where multi-photon microscopy is used to search for possible morphological asymmetries between the two brain sides. Moreover, we report on recently published results obtained by two-photon calcium imaging for functional mapping of the antennal lobe aimed at comparing patterns of activity evoked by different odours. Finally, possible links to the results of behavioural tests, measuring asymmetries in single-sided olfactory memory recall, are discussed.

**Introduction**

The brain of the honeybee (*Apis mellifera* L.) (Hymenoptera: Apidae), is an important model for small neural networks. In spite of the fact that it contains less than one million neurons (Menzel and Giurfa 2001), it shows a remarkably rich variety of cognitive performances. The olfactory pathway within the brain is of special interest to a large community of neuroscientists, since it provides for revealing experimental tests of information coding, learning, and memory models (Giurfa 2007). Over the years, various techniques have been developed to investigate the single odour processing steps from the periphery, the antennae, via the central nervous system (CNS), to the behavioural response.

Functional probing the peripheral olfactory system became possible after the development of the electroantennography (EAG) (Schneider 1957), which allowed for the first time to record the sum of the olfactory receptor responses to odour stimuli presented to the antenna. Detailed morphological information about peripheral olfactory sensitive structures, became instead accessible thanks to innovative high resolution imaging techniques like scanning electron microscopy (SEM), by which sensilla have been identified as carriers of odour receptors (Dietz and Humphreys 1971).

Early investigation of activity in the primary olfactory centres within the insect brain, the antennal lobes (ALs), was performed by means of extracellular recording (Boeckh 1974). In this respect, the major breakthrough was provided by the combination of intracellular recording with morphological imaging (Matsumoto and Hildebrand 1981), giving for the first time the ability to perform systematic analysis of specific neurons.

Pioneering work directed at optically imaging the antennal lobe activity in the honeybee brain were performed by Lieke in 1993, and thanks to the development of calcium sensitive dyes (Grynkiewicz et al. 1985) this soon became one of the most successful techniques for investigating the odour-evoked activity of the AL's functional subunits, the so called glomeruli (Galizia and Vetter 2004). Each glomerulus connected to a single olfactory receptor family were categorized in four classes, T1-T4, of which T1 is projecting via the lateral-axonal tract (l-ACT), T3-T4 via the median axonal tract (m-ACT) into mushroom body and lateral horn (Galizia et al. 1999b), the higher centre of the brain for multisensory integration. Here it undergoes further processing steps before the outcome of the olfactory processing network is finally projected out of the central nervous system. Motor neurons carry impulses to the muscles like those of the bee's proboscis. Thus responses to odour stimuli can be studied by the Proboscis Extension Reflex (PER) paradigm, an experimental procedure based on Pavlovian conditioning (Pavlov 1927). PER was for the first time experimentally conditioned in honeybees by Kuwabara in 1957 and has now become the most utilized procedure for exploring associative learning and memory recall in bees (Menzel and Müller 1996). In this paper we resume some experiments that combine the aforementioned techniques to investigate left-right asymmetries in the honeybee olfactory circuit.

The phenomenon of functional asymmetries was first reported in the human brain 150 years ago (Broca 1861). After that, lateralized behaviours have been discovered also in vertebrates in the 1970's (Nottebohm 1970; Rogers and Anson 1979), they were recognized by the time to be widespread among all taxonomic groups (Vallortigara and Rogers 2005; MacNeilage et al. 2009). But only recently, evidence of their presence in invertebrates was demonstrated (e.g. Kells and Goulson 2001; Ades and Ramires 2002; Hobert et al. 2002).

In the honeybee, side-specificity of odour learning was observed in behavioural studies (Sandoz and Menzel 2001; Letzkus et al. 2006; Rogers and Vallortigara 2008) as well as in morphological studies at the level of the antennae (Letzkus et al. 2006). On the contrary, from investigations of the primary olfactory centres in the brain so far only symmetric results were reported regarding morphology (Winnington et al. 1996) as well as the activity pattern (Galizia et al. 1998).

The approach of the experimental studies carried out by our research groups and summarized here, was to systematically trace asymmetries along the olfactory pathway using both established techniques as well as new imaging methods to acquire morpho-physiological, functional, and behavioural data. The focus of this review will be placed in particular on the experimental methods and their relative advantages and disadvantages.

**Scanning Electron Microscopy**

In Frasnelli et al.'s (2010a) anatomical differences in the number of sensilla between the right and the left antennae of the honeybee were investigated by SEM. Compared to Letzkus et al. (2006), the whole antenna was imaged, various types of sensilla where considered, and the study was based on a larger sample. The antennae were cut at the base of the scape and attached to a circular stub by double-sided conductive tape (TAAB Laboratories Equipment). All samples were gold-coated, guaranteeing electrical conductivity during imaging. Antennae were imaged using a XL 30 Field Emission Environmental Scanning Electron Microscope (FEI-Philips) in high-vacuum mode, from four different viewpoints, covering almost the complete surface (for details see Frasnelli et al. 2010a). The segments 3 to 9 were scanned longitudinally at a magnification of 600x, the 10th smaller apical segment with a magnification of 800x, segments 1 and 2 were omitted, since they don't carry olfactory receptors.

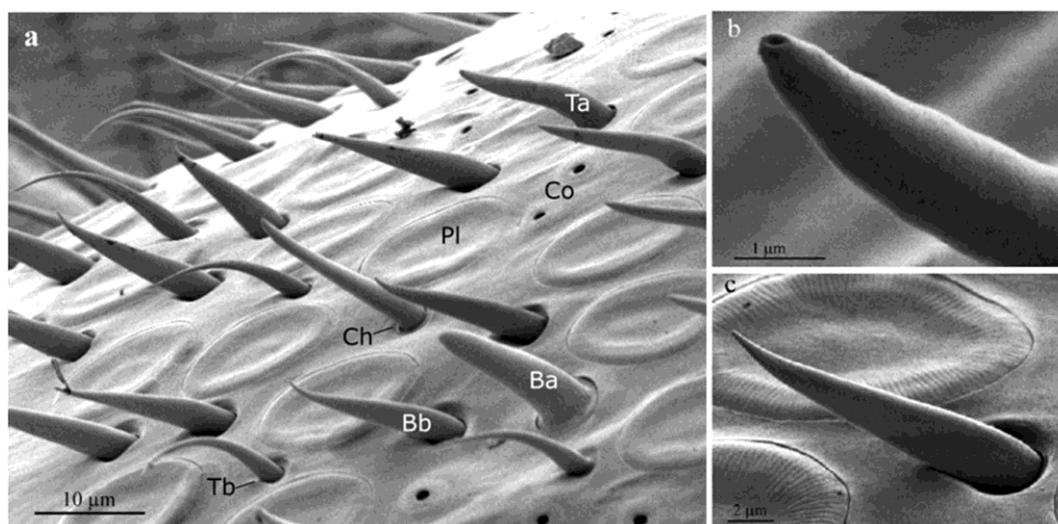

**Fig. 1** SEM image of the A. mellifera forager antenna: (a) dorsal view of a medial segment of the flagellum; (b) detail of a sensillum chaeticum; (c) detail of a sensillum basiconicum tapered. Labelled sensilla types in (a): Pl: sensillum placodeum; Ta: s. trichodeum type a; Tb: s. trichodeum type b; Ba: s. basiconicum thick; Bb: s. basiconicum tapered; Ch: s. chaeticum; Co: s. coeloconicum; (reprinted from Frasnelli et al. 2010a; Copyright (2011) with permission from Elsevier)

Eight sensillum types were tagged and counted on all acquired images by using image analysis software (UTHSCSA ImageTool Version 3.0). In particular, the putative olfactory sensilla were identified: s. placodea, s. trichodea type a and b, and s. basiconica thick and tapered;as well as the sensilla with non-olfactory functions: s. coeloconica, involved in perception of temperature, carbon dioxide and humidity, s. campaniformia, considered as hydro- and thermoreceptors and/or mechanoreceptors and s. chaetica, sensitive to mechanical or gustatory stimuli (Dietz and Humphreys 1971; Whitehead and Larsen 1976)

The SEM images of left and right antennae were analysed (*N*=14), comparing the number of olfactory and non-olfactory sensilla per antenna segment, the results are summarized in Figure 2. A significant asymmetry of the number of olfactory sensilla (Fig.2a) was detected throughout all segments in favour of the right antenna (ANOVA analysis of variance; $F_{(1,13)}$=5.56 *p*<0.05). For non-olfactory sensilla (Fig.2b), there was evidence of asymmetry in the segments 3–9, but in the opposite direction, with more non-olfactory sensilla on the left than on the right antenna (ANOVA analysis of variance; $F_{(1,13)}$=6.07 *p*<0.05).

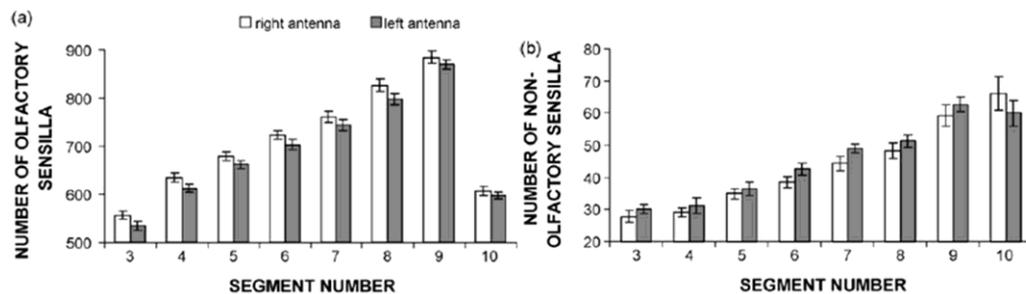

**Fig. 2**. Number of (a) olfactory and (b) non-olfactory sensilla per antenna segment in A. mellifera forager (N=14), obtained from antenna SEM images. Mean numbers of sensilla on the right (white bars) and left antennae (gray bars) are expressed with their standard errors. As clearly evident, the majority of the olfactory sensilla are located on the right antenna, while non-olfactory sensilla prevail instead on the left one (reprinted from Frasnelli et al. 2010a; Copyright (2011) with permission from Elsevier)

These data confirm and complete the results of Letzkus et al. (2006), suggesting that an asymmetry in the olfactory pathway may partially arise from a significant difference in number of olfactory sensilla in favour of the right antenna. However, as we discuss below, there is evidence of changes in the role of the left and right antenna during memory consolidation (Rogers and Vallortigara 2008) that cannot be explained by sensilla asymmetries only.

**Electroantennography**

Anfora et al. (2010) measured via electroantennography the overall depolarisation potential of the antenna's olfactory receptor neurons after odour presentation. The Experimental setup is depicted in Figure 3a: single antennae were cut at the level of the scape and brought in contact with the microcapillaries by means of two joystick micromanipulators (MN-151, NARISHIGE). Microcapillary glass electrodes were pulled to a tip size matching the antenna diameter of ~200μm (PP-83, NARISHIGE) and filled with Kaissling saline solution (Kaissling and Thorson, 1980). The signal was picked up by a high impedance probe followed by a fixed gain (5x) preamplifier and a high pass filter (INR-5, SYNTECH) suppressing 50Hz noise. The output was recorded by an acquisition module (IDAC-232, SYNTECH), connected to a computer, which was synchronized with a stimulus controller (CS-55, SYNTECH). For each subject, EAG peak responses (mV) (Fig.3b) were measured from both the right and the left antenna.

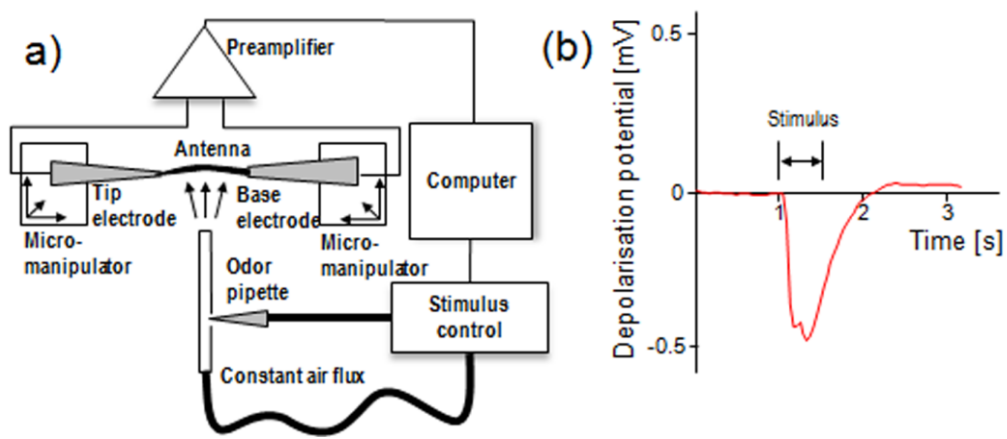

**Fig. 3** a) Schematics of the electroantennography experiment: electrophysiological recording setup synchronized with an odour stimulus generator. b) Typical depolarisation curve of an antenna exposed to a 500ms odour stimulus

Five samples of either isoamyl acetate (Sigma–Aldrich, >99.7% purity), component of the honeybee alarm pheromone (Bloch et al. 1962), or (-)-linalool (Sigma–Aldrich, >98.5% purity), a common floral odour, dissolved in hexane at concentrations ranging from $10^{-2}$ to $10^{2}$ µg/µl were prepared. Aliquots of the test solutions (25µl each), absorbed by pieces of filter paper (1cm$^2$), were inserted into individual Pasteur pipettes, while pure hexane and empty pipettes served as control. Odour puffs of 500ms, created by the stimulus controller, were injected into an air stream in ascending concentration order. The total air flux was kept constant at ~30ml/s to avoid mechanical stimuli. Odour puffs were emitted in 60s intervals. The results (Fig.4 and Anfora et al., 2010) revealed a similar pattern as the SEM experiments: the EAG responses, elicited by both tested compounds on the right antenna, were significantly higher than those on the left one (ANOVA: $F_{(1,15)}=5.12$; $p<0.05$). As expected, a significant increase in their EAG responses with increasing doses of isoamyl acetate and (-)-linalool was observed, with the latter showing saturation for the highest concentration.

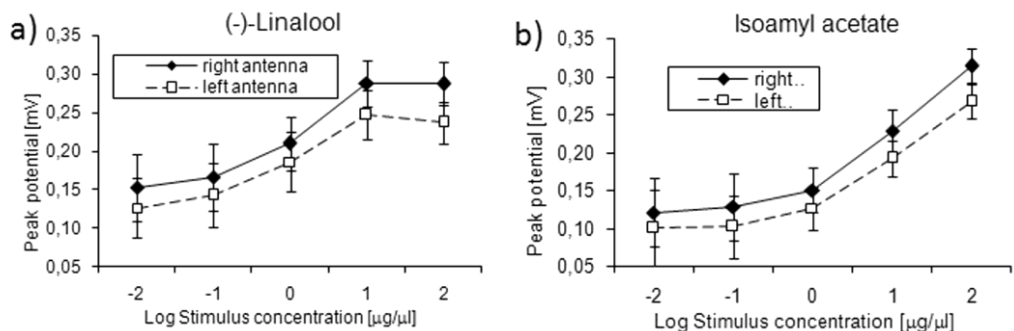

**Fig. 4** Electroantennography results: Mean response potential ± Standard Errors of right and left antenna of *A. mellifera* foragers (*N*=16) to (a) (-)-linalool and (b) isoamyl acetate measured over five orders of magnitude in concentration. Empty squares and black rhombs, representing responses of the left and right antenna, respectively, confirm an enhanced response of the right antenna for both odours (reprinted from Anfora et al. 2010; Copyright (2011) with permission from Elsevier)

The electrophysiological measurements of Anfora et al. 2010 on single antennae confirm the morphological studies discussed above (Letzkus et al. 2006; Frasnelli et al. 2010a), showing the presence of a significant response asymmetry favouring the right antenna (Anfora et al. 2010).

However, as noted above, behavioural experiments evidenced changes in lateralization associated with memory consolidation (Rogers and Vallortigara 2008) with a dominance of the right antenna at short timescales and a dominance of the left antenna at long (>3h) time-scales. This suggests a contribution of the CNS to lateralization, motivating new research efforts aimed at looking for lateralization in the primary olfactory centres of the honeybee brain, the antennal lobes.

**Two-photon microscopy**

Two-photon-microscopy is a well established imaging technique which offers unprecedented optical sectioning capabilities thanks to its enhanced penetration depth and axial resolution (Denk et al. 1990). Haase et al. (2011) introduced this technique to the field of honeybee neuroimaging through the implementation of an in-vivo imaging platform allowing to obtain both morphological and functional data from the whole antennal lobe of the honeybee. This imaging tool provided the capability to conduct studies aimed at finding the presence of volumetric lateralization of the glomeruli, as well as analyzing the glomerular activity for asymmetries in their odour response maps. Given the novelty of these procedures, they will be described here in some details.

**Animal Preparation**

For the two-photon microscopy experiments reported here (Haase et al. 2011; Rigosi et al. 2011), bees were prepared according to a well-established protocol (Galizia and Vetter 2004). After chilling to immobility, bees were fixed to a custom made imaging stage using dental waxes (Kerr; Siladent). A window was cut into the head's cuticula above antennal lobes and mushroom body and glands and trachea were gently removed. A pond for the objective immersion liquid was formed above the imaging region using plastic cover slip, sealed with silicon (Kwik-Sil, WPI). This protocol was followed closely which should also guaranty a comparability of the imaging results with previous experiments.

**Morphological imaging**

In Rigosi et al. (2011) bees' ALs were morphologically stained with a membrane-selective dye. First, the neural sheath was digested with a 1% solution of Protease Type XIV (Sigma-Aldrich). Then the bee brain was homogeneously stained by bath-application of the cell-permeant dye RH795 (50μM, Invitrogen).This preparation method was chosen by the authors as the least invasive technique, avoiding artefacts due to brain extraction, fixation, and dehydration (Bucher et al. 2000).

The imaging setup is sketched in Figure 5a. It consists of a two-photon microscope (Ultima IV, Prairie Technologies) combined with an ultra-short pulsed laser (Mai Tai Deep See HP, Spectra-Physics). The excitation wavelength was 1040nm for two-photon excitation of RH795, detection was centred around 607nm. The beam is strongly focussed onto the sample by a water immersion objective (40x, NA 0.8, Olympus).

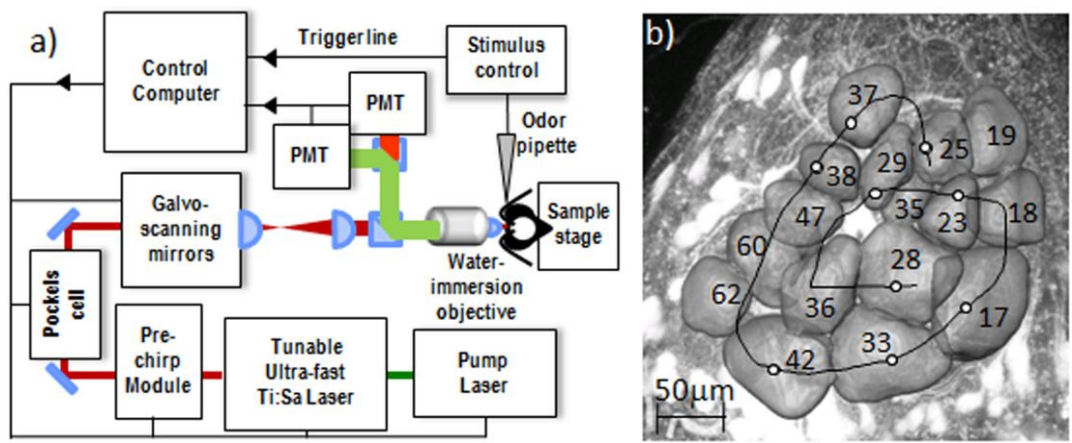

**Fig. 5** a) Schematic setup of the two-photon microscope: A tunable ultra-short pulsed laser (Mai Tai Deep See HP, Spectra-Physics) is dispersion-compensated in pre-chirp module. A Pockels cell controls the light intensity and galvo-mirrors allow for fast and variable scanning. The beam is strongly focussed onto the sample by a water immersion objective (40x, NA 0.8, Olympus). Fluorescence is collected by the same objective, separated from the backscattered excitation light by a dichroic beam-splitter, split into green and red detection channels by dichroics and band-pass filters (Chroma Technology), and detected by Photomultiplier tubes (PMT, Hamamatsu Photonics). A computer controls all microscope parameters and synchronizes imaging with a odour stimulus generator. (b) Axial projection view of a right AL in *A. mellifera* forager, reconstructed volume images of a subset of T1 glomeruli (labelled according to Galizia et al. 1999b), the black spiral represents the custom defined scan-trace for fast functional imaging of the glomerular activity (reprinted from Haase et al. 2011 with permission from OSA)

A point-spread-function measurement verified the microscope's resolution to be diffraction limited to a Gaussian width $\sigma_{x,y}$=230nm in the plane and $\sigma_z$ = 1.1µm axially (for the 40x lens at 800nm excitation wavelength). For later volume reconstruction, image stacks of 3µm axial distance were taken. Dye diffusion limited the imaging depth to ∼150µm. Post-processing for 3D reconstruction, image segmentation, and volumetric measurements was performed using the software Amira (Visage Imaging). As a test sample five primary glomeruli were chosen in the study by Rigosi et al. (2011) which allow to be easily identified and exhibit very diverse response profiles (Galizia et al. 1999a; Haase et al. 2011). The latter would cover the possibility of an odour-specific lateralization (see Rigosi et al., 2011).

The volume measurements of the reconstructed glomeruli gave results about twice the size of those from dehydrated samples investigated in previous studies (e.g. Winnington et al. 1996; Hourcade et al. 2009). This confirmed that impressively strong shrinkage effects were avoided by this imaging method.

The determined relative volume imbalance between left ($V_L$) and right ($V_R$) ALs was quantified by the lateralization index $L=V_R/(V_R+V_L)$. The results reported by Rigosi et al. (2011) showed no evidence of significant morphological asymmetry between sides, only the expected significant difference in volume among different glomerular classes was observed.

Thus, this experiment did not only confirm findings of Winnington et al. (1996), the only previously published study comparing AL volumes between the two sides of the brain, but it strengthened these results, improving the axial resolution of the AL images by almost an order of magnitude with respect to the one by Winnington et al., which was based on histological slices taken at distances of 25µm. Moreover, possible artefacts due to anisometric shrinkage and diffraction-index mismatch (Bucher et al. 2000) could be avoided, representing a clear advantage over the common method in morphological imaging of the bee brain, which implied extraction, fixation, and clearing of the brain sample (e.g. Galizia et al. 1999b; Hourcade et al. 2009),

The findings of Rigosi et al. (2011) do not necessarily conflict with the morphological asymmetries found in the peripheral part of the olfactory pathway, the antennae, since there are several possible effects that might superimpose or compensate for this peripheral imbalance. Firstly, there is no linear correlation of sensilla number to glomerular volume, since the single glomeruli represent a convergent site for all odour receptor neurons of one type (Galizia and Szyszka 2008) and even within a receptor class the number and type of neurons per sensillum can vary drastically (Kelber et al. 2006). So the peripheral lateralization might cause only minor asymmetries between the AL, too small to be distinguished from statistical variations. Besides that, there are reports about enormous odour-experience dependent volumetric plasticity in specific glomeruli (Winnington et al., 1996; Sigg et al., 1997; Hourcade et al. 2009), which again might cause statistical variations, overshadowing a minor lateralization effect, since the odour experience of the bees imaged by Rigosi et al. (2011) could not be controlled. In future morphological studies, efforts have to be made to control the subjects' odour experience, e.g. by uniform conditioning before the experiments. Moreover, in order to look for an effect corresponding to the lateral shift from short-term memory (STM) to long-term memory (LTM) in the antennal lobe, the time span between conditioning and imaging should be varied systematically.

**Functional imaging**

Apart from morphological measurements, first prove-of-principle experiments in functional imaging using two-photon microscopy have been reported by Haase et al. in 2011, aimed to further decipher olfactory code, to study the principles of odour learning and memory, and to search for possible asymmetries in the functional signals along olfactory pathway.

For in-vivo functional imaging, the projection neurons where selectively stained by the backfill method (Gelperin and Flores 1997) using a solution of the calcium sensitive dye

fura2-dextran (Invitrogen) and 2% Bovine Serum Albumin (Sigma-Aldrich) For dye injection, glass capillaries (ID 1.15mm, OD 1.65mm, Garner Glass Comp.) where pulled to tip sizes of ~5µm (PP-83, NARISHIGE), coated by rolling them in the viscous dye solution, then tips were broken, leaving a final tip diameter of ~20µm. Dye-coated micro-tips were carefully injected into the antenno-cerebralis tracts below the α-lobe. Then the cuticula was reclosed and the animals were stored for 20h in a dark, cool, and humid place in order for the dye to diffuse into the AL. Before the imaging session, the cuticula, the glands, and the trachea above the AL were removed and a pond was formed and filled with Ringer's solution (Galizia and Vetter 2004).

The imaging setup in Haase et al. (2011) was similar to that for morphological imaging described above, except for the excitation wavelength of 800nm for resonant two-photon excitation of the free fluorophore fura-2, the detection channel which was filtered by a 70nm bandpass around 525nm, and for the time-optimized laser scanning technique. To record the spatio-temporal response pattern with the highest possible temporal resolution, the excitation laser was scanned within an arbitrary horizontal plane along one-dimensional custom-defined trace (Fig.5b). These line scan traces were chosen to cross all glomeruli of interest and allowed scanning frequencies as high as 70Hz. An optimal signal-to-noise ratio was achieved with laser powers of about 10mW at the sample surface, without observing any induced photo-damage during the measurement. The temperature of the experimental environment was stabilized to 29°C (Franke 2009).

Functional imaging was performed measuring the intra-neuronal calcium concentration, which increases with enhanced neuronal activity. When the $Ca^{2+}$-sensitive fluorophore binds calcium, its absorption spectrum is shifted away from the laser excitation wavelength. This causes a drop in the measured fluorescence intensity, manifesting themselves in dark bands in the scanline-over-time maps at the positions of the active glomeruli (Fig.6).

We see changes in the staining quality between single subjects causing variations in the absolute fluorescence intensity, but, since fura-2 is ratiometric Calcium- marker, the relative shift due to the $Ca^{2+}$ concentration change remains constant. This was confirmed in our experiments..

A stimulus controller (CS-55, SYNTECH) delivered odour stimuli at constant total air flux (30ml/s). The odour puffs came from Pasteur pipettes in which 10µl of an odour solution (1:10 in mineral oil) were deposited on filter paper. All command signals and acquisitions were controlled by a common gate, which allowed precise synchronization of the involved pulses.

The experimental cycle began by starting the image acquisition. After 3s the stimulus generator received a trigger releasing an odour puff of 2s length. The exact arrival time of the odour at the bee antenna was measured and found to be stable within 10ms, which allowed for accurate measurements of the neuronal response delay. These measurements were performed by optically monitoring a cantilever at the position of the antenna before the functional imaging. To assess concentration fluctuation during the odour response measurements, a photo-ionization detector would have to be added to the setup. After 9s image acquisition stopped and data evaluation was automatically performed via Matlab (Mathworks) scripts.

All acquired data were corrected for photo-bleaching, but this turned out to be negligible, as the fitted exponentials were mostly flat. A 2D running average filtering was used to reduce the noise level. Spatial averaging was performed over 30 pixels corresponding to 17.4 $\mu$m (the minimum glomerular size). Temporal averaging was applied over 5 pixels corresponding to 88 ms preserving all main dynamic features of the data. The spatio-temporal response pattern of each imaging experiment was then analysed. Thanks to a reduced photo-damage of two-photon microscopy (Denk and Svoboda1997) due to the limited absorption volume in the sample, photo-bleaching was strongly suppressed, allowing for repeated recordings. To measure the total response intensity , each odour was applied three times consecutively and data-sets were averaged. This improved the signal-to-noise ratio, which became especially important at greater imaging depths where random fluctuations increased.

The reduced photo-damage allowed also to extend the imaging sessions up to 5h before an essential drop in the brain activity could be noticed.

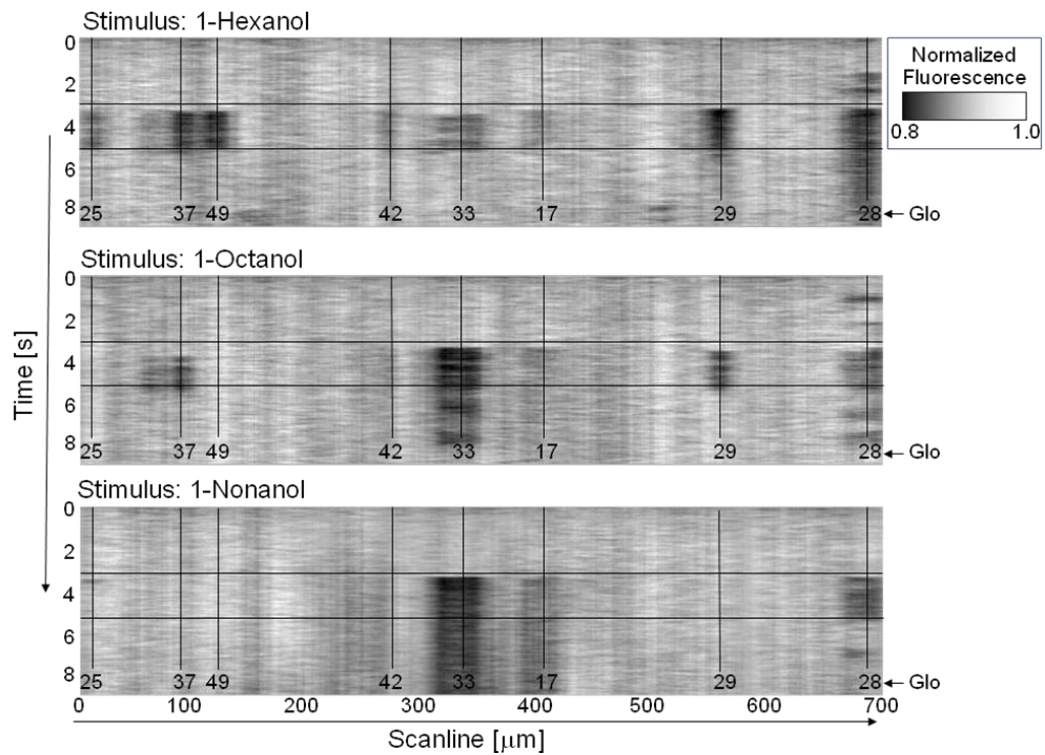

**Fig.6** Calcium response maps for three different odours stimuli:(a) 1-hexanol, (b) 1-octanol, (c) 1-nonanol, recorded along the scanning trace in Figure 5b. The stimulus period is enclosed in horizontal lines, responding glomeruli centres are marked by vertical lines, numbers label the identified T1 glomeruli according to Galizia et al. 1999b (reprinted from Haase et al. 2011, with permission from OSA)

The resulting spatio-temporal response patterns of the AL are presented in Figure 6. The calcium concentration changes were recorded along the line traces indicated in Figure 5b. To compare the performance of two-photon microscopy with conventional methods, the applied odour stimuli were chosen to be three well studied plant volatiles: 1-hexanol, 1-octanol, and 1-nonanol (Galizia et al. 1999a; Peele et al. 2006). The odour response signals detected by Haase et al. (2011) went up to 20% direct fluorescence intensity change (Fig.7), which represents more than a fourfold improvement with respect to comparable experiments using full-field imaging. The recorded odour response maps (Fig. 6) reproduced features which had already been observed by conventional single-photon fluorescence microscopy, such as the very strong response of glomerulus T1-28 to all tested odours decreasing with increasing odour's carbon chain length, and those of T1-33 increasing with the odour's carbon chain length (Sachse et al. 1999; Peele et al. 2006). Likewise, 1-hexanol was found to produce the broadest response spectrum of the tested odours. Strikingly different from previously published data obtained with full-field microscopy (Peele et al. 2006) were the quite strong responses of glomeruli T1-29 and T1-37 for both 1-hexanol and 1-octanol, which might be due to the enhanced sensitivity of the imaging system.

As in wide-field microscopy experiments (e.g. Galizia et al. 1999) we see a variance in the total responses strength to the same odour among different preparations. We attribute this mainly to differences in age and odour experience, which we were unable to control in our outdoor hives. For an improved quantitative analysis, odour conditioning before the imaging might be a solution.

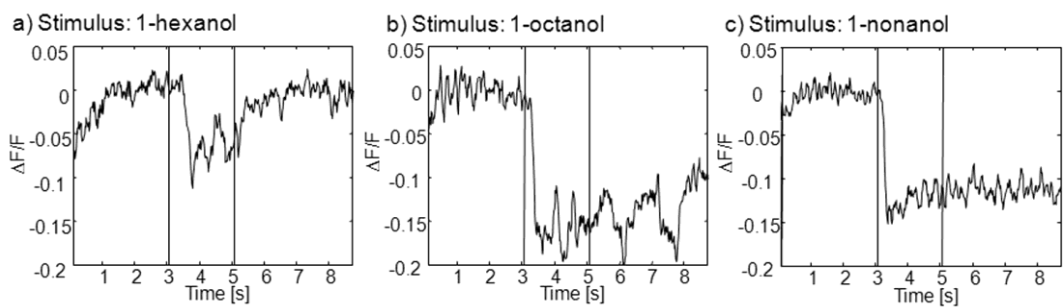

**Fig. 7** Temporal activity pattern of the single glomerulus T1-33 (averaged 17,4 μm around the centre, compare Fig.6). The relative change in fluorescence is plotted as a function of time, the stimulus period is enclosed by vertical lines. Besides changing response latencies, also slow temporal variations of the response signal can be observed. Both might be part of a complex odour code (reprinted from Haase et al. 2011 with permission from OSA)

The high temporal resolution of the study is demonstrated in Figure 7, where response curves of glomerulus T1-33 to all tested odours are shown at an imaging depth of 25μm.

The time resolution and the precise stimulus synchronization, predestine this method for a systematic investigation of response delays and their dependence on the odour stimuli, as well as the study of a possible specificity of the observed temporal variations of the response signal. Both response latency (Müller et al. 2002) and synchronized oscillations (Laurent 2002), previously observed in electrophysiological measurements, are hypothesized to be part of a general odour code. The larger penetration depth due to infrared excitation together with the backfill staining method allowed Haase et al. to optically access the whole antennal lobe. Figure 8a shows an image stack down to 270µm, where the deepest glomeruli are located.

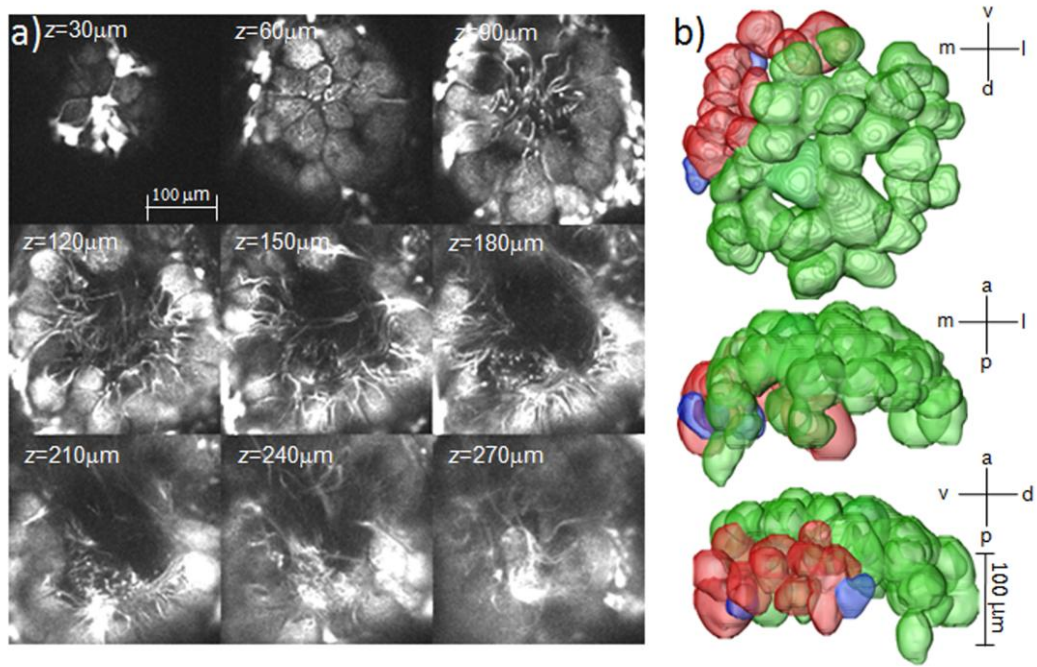

**Fig.8** a) Image stack of a right AL in *A. mellifera* down to 270 µm. b) Reconstructed volume images of a subset of glomeruli. Green colour marking the T1 glomerular class, red the T2, and blue the T3 glomerular class(reprinted from Haase et al. 2011 with permission from OSA)

A volumetric reconstruction of single glomeruli was performed down to a depth of 200µm, the limit to which they were so far able to identify glomeruli with certainty. Figure 8b shows volume images of these glomeruli along all principle axes. The reconstructed glomeruli have the following colour code: green ones belong to the T1 sensory group projecting into the l-ACT axonal tract, while blue and red coloured ones belong to the deeper laying T2 and T3 groups, respectively, both projecting into the m-ACT.

These results manifest how two-photon microscopy led to improvements in honeybee neuroimaging in five respects: The enhanced signal amplitude of 20% fluorescence change together with the increased temporal resolution might allow to resolve features that have so far been hidden to conventional full-field microscopy, e.g. possible lateralization in the activity pattern, which could not be revealed in a previous study (Galizia et al. 1998). Increased penetration depth allowed optical access also to profound regions within the AL, which so far could not be systematically investigated. Higher axial resolution will allow to extend the odour response maps (Galizia et al. 1999a) into a third spatial dimension, including deeper lying glomerular classes (T2-T4). This is of great interest since intracellular recording (Müller et al. 2002; Krofczik et al. 2009) and imaging of their axon terminal boutons in the mushroom body (Yamagata et al. 2009) predicted a fundamentally different role for this glomerular classes in olfactory coding and memory. Lateralization effects in short- and long-term memory recall, observed in behavioural studies (Rogers and Vallortigara 2008) might be caused by plasticity in these regions. Furthermore, for the first time a convenient temporal resolution could be achieved by an optical method, giving hope to add a fourth, temporal dimension to the glomerular response maps. The temporally resolved activity pattern will allow investigating the role of response latencies, already observed by intracellular recording (Müller et al. 2002), as a possible odour coding parameter, which was confirmed in other species (Junek et al. 2010). The observed temporal variations of the response signal should be checked for specificity, which would support dynamic odour coding theories (Laurent 2002). So there is great potential to verify whether several proposed coding models apply to the bee's olfactory pathway and whether a possible asymmetry that could not be detected in the morphology of T1 glomeruli, might be found in these complex codes. The fundamental question whether odour learning manifests itself in functional plasticity within ALs has not been answered satisfyingly yet, since a series of experiments came to very different results. Various studies (Faber et al. 1999; Sandoz et al. 2003; Fernandez 2009) have found plasticity in the calcium response signals of the olfactory receptor neurons (ORNs), while others could not confirm this effect neither in the ORNs (Hourcade et al. 2009) nor in the projection neurons (Peele et al. 2006; Roussel et al. 2010). Possible explanations for these differing results are different time intervals between odour conditioning and imaging, which should be investigated systematically.  This could be realised thanks to the reduced photo-damage of two photon microscopy, which permitted Haase at al. to extend imaging sessions up to 5h, which may allow in the future to study plasticity in the AL in real time over extended periods. The latter would also be extremely interesting regarding the search for an effect in the AL, corresponding to the lateral shift from the right to the left brain side, which was observed in behavioural experiments between the $1^{st}$ and the $6^{st}$ hour after odour conditioning (Rogers and Vallortigara 2008) and was hypothesised to be associated with memory consolidation from STM to LTM. But this experiment will be quite ambitious since the restriction that in one samples only one AL can be prepared requires a big statistical basis to detect lateralization on top of the response fluctuations between different individuals. Nevertheless, considering the few knowledge about physiological correlates of lateralized behaviour  it might be a challenging and intriguing question to address. In honeybees, in fact, the only evidence at the central level about lateralization it has been recently revealed; Biswas and colleagues in fact demonstrated a difference between sides in the distribution of neuroligin-1 a protein involved in learning and memory when honeybees were only left or right antennae amputated (Biswas et al., 2010)

**Behavioural experiments**

The last stage of the olfactory pathway, the behavioural response, was analysed in Rigosi et al. (2011). The experiments were linked to the morphological imaging study by choosing odour stimuli, that strongly involved the imaged glomerular subset as shown in previous recorded response maps (Galizia et al.1999a, Peele et al. 2006; Haase et al. 2011), searching for a possible odour dependence of lateralization.

Rigosi et al. (2011) cooled bees until immobility and mounted them into customized holders (Bitterman et al 1983; Rogers and Vallortigara 2008). The subjects were assigned randomly to groups for the occlusion of one antenna. The bees in one group had their left antenna coated with silicone (Silagum-Mono, DMG), those in the second group had their right antenna coated, while both the antennae of the bees belonging to the third group were left uncoated. One hour later each bees groups was subdivided into three groups to be trained to three different odours, involving odour compounds from floral parts like (-)-linalool (Sigma-Aldrich, purity >98.5%), but also volatiles released by green organs of the plants like 1-octanol and 2-octanone (both Fluka, purity >95%). These odours served as a positive conditioned stimulus (CS+) together with 1M sucrose solution as a food reward (unconditioned stimulus, US). 10µl of each odour compound was dissolved in 3ml of the sucrose solution. A saturated saline solution served as negative conditioned stimulus (CS-).

Three trials spaced 6min apart were performed, where a droplet of the odour sugar solution at the end of a 23 gauge needle was held 1cm above the bee's antennae. After 5s the antennae were touched, which led to PER. The bee was then allowed to ingest the drop of the CS+/US solution. The procedure was repeated with the saline solution, which did not trigger PER but avoidance by moving the antennae away from the droplet. STM recall was tested 1h later by presenting the odour dissolved in distilled water or the saline solution and holding the droplet 1cm above the antennae without touching it. These CS+ and CS- stimuli were presented for 5s. Each bee was tested a total of 10 paired trials, the CS+ and CS- were given in a random order with 60s between each odour presentation. The PER responses were recorded and the percentage of success was scored as extensions of the proboscis to odours and no extension to saturated salt solution [(*PER*(CS+)-*PER*(CS-)]/(#paired trials).

The results of the conditioning and retrieval tests by Rigosi et al. (2011) can be summarised as follows: the right antenna seemed to show better recall than the left antenna in odour STM retrieval, with a strong odour dependence. However this effect was strongly modulated by the odour type. While the authors revealed no significant lateralization for 1-octanol (Kruskal-Wallis test; $X^2$=1.02; *p*=0.6, *N*=70) and 2-octanone (Kruskal-Wallis test; $X^2$=0.97; $p$=0.62, $N$=61), bees trained with (-)-linalool showed a significant effect of the antenna in use (Kruskal-Wallis test; $X^2$=9.91; $p$<0.01, $N$=63) favouring the right antenna as was previously demonstrated elsewhere (Frasnelli et al. 2010a).

These results confirm that olfactory learning asymmetry in bees is indeed odour-dependent. In previous studies, documenting lateralization (Letzkus et al. 2006; Rogers and Vallortigara 2008; Frasnelli et al. 2010a; Frasnelli et al. 2010b; Anfora et al. 2010), tested odours were pheromones or volatiles from plant's blooms, all of great importance to forager bees. In the experiments of Rigosi et al. (2011) also unspecific compounds (1-octanol, and 2-octanone), released by the green organs of the plants and being of minor importance in pollinator plant interaction, were tested. Those showed no significant olfactory lateralization effect. While (-)-linalool, one of the most common derivate of floral scents, which was shown to be a key compound in the identification of odour mixtures (Reinhardt et al. 2010), produced significant olfactory lateralization, confirming the findings of Anfora et al. (2010). This suggests that the occurrence of lateralization might depend on the odours' importance to pollinators, which means lateralization of olfactory information might be advantageous only if the odour is biologically relevant to the bees. Being specialized just in one side for specific odour memory might represent an important advantage in terms of leaving the other hemisphere free to perform a different parallel task (for an example in vertebrates see Rogers et al. 2004). Moreover, this difference between sides in naive individuals apparently does not impair sensory sensitivity as for instance olfactory asymmetries has been showed in odour tracking in *Drosophila melanogaster* (Duistermars et al. 2009).

This assumption has to be confirmed by further experiments using odour compounds of different origin and significance for honeybees and by repeating the memory retrieval tests at different time points, ruling out that the experiments were accidentally performed at the point where the lateral shift from STM to LTM temporarily causes symmetric response efficiencies.

This aspect of possible distinct memory mechanisms for more or less relevant odours is extremely interesting also for future optical imaging experiments, where the search for possible lateralization could be focused on glomerular subsets with narrow response spectra to the relevant odour classes.

## Acknowledgments

This work has been realized also thanks to the support from the Provincia Autonoma di Trento and the Fondazione Cassa di Risparmio di Trento e Rovereto. C.V. acknowledges Provincia Autonoma di Trento (project COMFI) and National Institutes of Health grant RO1EB006432.